\documentclass[aps,twocolumn,superscriptaddress,showpacs]{revtex4}
\usepackage{amsmath,graphicx,multirow}
\newcommand{\half}{\mbox{$\textstyle \frac{1}{2}$}}
\newcommand{\ket}[1]{| #1 \rangle}
\newcommand{\bra}[1]{\langle #1 |}
\begin{document}
\title{The effects of reduced ``free will" on Bell-based randomness expansion}

\author{Dax Enshan Koh}
\affiliation{Centre for Quantum Technologies, National University of Singapore, 3 Science Drive 2, Singapore 117543}
\author{Michael J.W. Hall}
\affiliation{Centre for Quantum Computation and Communication Technology (Australian Research Council), Centre for Quantum Dynamics, Griffith University, Brisbane, QLD 4111, Australia}
\author{Setiawan}
\affiliation{Department of Physics, National University of Singapore, 2 Science Drive 3, Singapore 117542}
\author{ James E. Pope}
\affiliation{Mathematical Institute, University of Oxford, 24-29 St Giles', OX1 3LB, UK}
\author{Chiara Marletto}
\affiliation{Mathematical Institute, University of Oxford, 24-29 St Giles', OX1 3LB, UK}
\author{Alastair Kay}
\affiliation{Centre for Quantum Technologies, National University of Singapore, 3 Science Drive 2, Singapore 117543}
\affiliation{Keble College, Parks Road, Oxford, OX1 3PG, UK}
\author{Valerio Scarani}
\affiliation{Centre for Quantum Technologies, National University of Singapore, 3 Science Drive 2, Singapore 117543}
\affiliation{Department of Physics, National University of Singapore, 2 Science Drive 3, Singapore 117542}
\author{Artur Ekert}
\affiliation{Centre for Quantum Technologies, National University of Singapore, 3 Science Drive 2, Singapore 117543}
\affiliation{Mathematical Institute, University of Oxford, 24-29 St Giles', OX1 3LB, UK}


\begin{abstract}
With the advent of quantum information, the violation of a Bell inequality is used as evidence of the absence of an eavesdropper in cryptographic scenarios such as key distribution and randomness expansion. One of the key assumptions of Bell's Theorem is the existence of experimental ``free will", meaning that measurement settings can be chosen at random and independently by each party. The relaxation of this assumption potentially shifts the balance of power towards an eavesdropper. We consider a no-signalling model with reduced ``free will" and bound the adversary's capabilities in the task of randomness expansion. \end{abstract} \maketitle

\newtheorem{theorem}{Theorem}

%\section{Introduction}
\textit{Introduction.---}A source of random data that can be trusted to be truly random, and not just repeating a pre-determined, apparently random, sequence is a vital resource in many applications. A real-world scenario is online gambling, where accusations abound of sites rigging the deck, etc. Quantum mechanics has long been known to provide intrinsic randomness, but it has recently been noticed that Bell tests allow us to go further: they provide \textit{quantitative} bounds for the amount of randomness that is generated \cite{3,roger,zurich}. Moreover, these bounds are device-independent, in the sense that they are obtained only from the observed statistics, without reference to a description of the physical system or the implemented operations. Two different types of bound can be achieved, either by assuming the validity of quantum mechanics or merely with the weaker assumption of no-signalling in a fully black-box scenario.

In a randomness expansion protocol, a pre-established stock of randomness (for instance, a string of random bits) is used to make measurement selections in a series of Bell tests, operated by two parties (Alice and Bob) in distantly separated parts of the same laboratory. The correlation statistics of the outcomes are used to violate a Bell inequality, giving a quantitative bound on the degree to which an adversary or eavesdropper (Eve) is excluded. This bound can be used to measure the randomness of the outcomes, which can be added to the stock of private randomness \cite{1}. To certify the private randomness produced, it is crucial to not only determine what we call the \textit{guessing probability} $G$ (defined below), but also to ensure that Eve cannot somehow fake this bound, perhaps by bypassing some of the assumptions used in the derivation of the bound. One of these assumptions is that Alice and Bob can randomly and independently select their measurements. While Alice and Bob could rely on making these choices with their own free will, in practice they use random number generators (RNGs), which Eve could potentially manipulate to deliberately introduce patterns undetected by standard statistical tests, giving rise to the interpretation that Eve compromises the free will of Alice and Bob.

We study the extent to which Eve, by influencing those measurement choices, can manipulate the degree of violation ($S$) of a Bell test using a no-signalling model \cite{2,5}. Eve does her best to pre-program the outcomes of Alice's and Bob's measurements so that, for prescribed $S$ and degree of influence upon Alice's and Bob's measurement choices, her probability of guessing the measurement outcomes correctly is maximized. The more influence she has, the less ``free will" can be attributed to Alice and Bob, and if they wrongly assume that they have complete free will, they can be fooled into thinking that their observed outputs are not predetermined.

Previous discussions of the free will assumption have quantified the concept in differing ways \cite{5,10,6}. The upshot is that free will seems to be a critical resource for the violation of Bell inequalities in order to derive their usual interpretation. Indeed, if free will is given up on 41\% of the runs of an experiment, singlet state correlations can be reproduced from classical correlations \cite{footnote}. 

An operational way of quantifying randomness involves the notion of \textit{guessing probability} or {\em predictability}: a process has large randomness if it is hard to guess its outcomes.
Here we establish bounds on the average probability of guessing an outcome of a Bell test, for a given amount of free will, using a variant of Hall's relaxed Bell inequalities \cite{7}. While these results require only the no-signalling restriction, for comparison, we also establish the limiting strategy for a quantum Eve who eavesdrops each run independently.

%\section{Overview}

\textit{Model.---}We work in the simplest scenario of two parties, each with two inputs and two outputs, for which the Clauser-Horne-Shimony-Holt (CHSH) inequality \cite{8} is the unique Bell test. The devices that Alice and Bob use are treated as black boxes, potentially prepared by Eve. The inputs are labelled $A_j$ and $B_k$ respectively, where $j,k \in  \{0,1\}$, and the outputs are labelled $a,b \in \{0,1\}$. The CHSH test is repeated a large number of times, yielding a probability distribution of the outputs $\{\tilde{p}(a,b|A_j, B_k)\}$, which we assume to be no-signalling. In terms of these probabilities, the CHSH correlation function $S$ can be defined as
\begin{equation}
S = \Big| \sum_{a,b,j,k \in \{0,1\}} (-1)^{a+b+jk} \tilde{p}(a,b|A_j,B_k) \Big|.
\end{equation}

In order to study eavesdropping strategies, we will impose that the probability of each input is equally likely, i.e.\ $p(A_j, B_k)  = \frac{1}{4}$ for all $j,k \in \{0,1\}$, which means that Alice and Bob, with no knowledge of the underlying strategy, are not able to detect any deviations of these probabilities from the uniform distribution $(\frac{1}{4}, \frac{1}{4}, \frac{1}{4}, \frac{1}{4})$ that they expect. Eve's control over the inputs and outputs is described by an underlying variable $\lambda$, corresponding to conditional probability densities $\tilde{p}(a,b|A_j, B_k, \lambda)$ and $\rho(\lambda| A_j, B_k)$. These are related by Bayes' theorem: $\tilde{p} (a,b|A_j, B_k) =  \int d \lambda \ \tilde{p} (a,b|A_j, B_k, \lambda) \rho(\lambda|A_j,B_k)$. The summation over $b$ and $a$ respectively produce the marginals $\tilde{p}^{(A)}(a|A_j, \lambda)$ and $\tilde{p}^{(B)}(b|B_k, \lambda)$. Note that the no-signalling assumption imposes that the marginal probabilities $\tilde{p}^{(A)}$ and $\tilde{p}^{(B)}$ are independent of $B_k$ and $A_j$, respectively.

\textit{Guessing probability.---}The guessing probability, or predictability, $G(\lambda)$ for a given underlying variable $\lambda$ is the maximum over all these marginal probabilities
$$
G(\lambda)=\max_{a,A_j,b,B_k}\left(\tilde{p}^{(A)}(a|A_j, \lambda),\tilde{p}^{(B)}(b|B_k, \lambda)\right),
$$
i.e.\ it upper bounds the probability of Eve guessing one of Alice's or Bob's outcomes, knowing a given $\lambda$ is being used on a specific run of the experiment. The guessing probability, for Alice, Bob or any observer without access to the underlying variables, is the weighted average of $G(\lambda)$ over $\lambda$, i.e.
\begin{equation}
G = \int d\lambda \ \rho(\lambda) G(\lambda),
\end{equation}
where $\rho(\lambda)$ is the probability distribution of the variable $\lambda$. Note that $G$ takes values in the closed interval $[\frac{1}{2},1]$, where $G= \frac{1}{2}$ ($G=1$) means that the underlying model is completely indeterministic (deterministic).

For a given Bell violation, tight bounds for $G$ have been calculated in the literature \cite{4} for the case of complete free will. In order to formulate the relaxation of free will, we define a free will parameter, $P$, as the maximum probability that a particular pair of measurement settings is chosen, maximized over all control variables $\lambda$, i.e. 
\begin{equation}
P = \max_{j,k, \lambda} p(A_j, B_k | \lambda). \label{P}
\end{equation}
This quantifies the maximum deviation of $p (A_j ,B_k | \lambda)$ from the uniform distribution, i.e.\ the extent of Eve's influence over the supposedly free choice. For a 2-party, 2-setting protocol, $P$ takes values in the interval $[\frac{1}{4}, 1]$; $P=\frac{1}{4}$ corresponds to the case of complete free will, while $P=1$ corresponds to a deterministic selection specified by Eve. This definition relates directly to the probability that a pair of inputs is chosen for a given underlying variable. While being more natural for our model, this differs from that given in \cite{2}, which involves conditional probability distributions of the underlying variable given the measurement inputs. Nevertheless, a correspondence between the two can be found via Bayes' Theorem. From these definitions, we obtain the following theorem (proved in Appendix \ref{sec:proof}):

\begin{theorem}
The maximum possible CHSH expectation value $S^{\max}(G,P)$, for a guessing probability $G$ and free will parameter $P$, for any no-signalling model with $p(A_j, B_k) = \frac{1}{4}$ (i.e. all inputs are equally likely), is

\begin{equation}
S^{\max}(G,P) = \begin{cases} 4 - 8(2G-1)(1-3P) & P \leq \frac{1}{3}, \\ 4 & P \geq \frac{1}{3}. \end{cases} \label{eqthm1}
\end{equation}

\label{thm1}
\end{theorem}
We illustrate this result with three limiting cases. If Eve knows exactly, for each instance of the measurement, what will be measured, then Alice and Bob have no ``free will" ($P=1$); their actions are predetermined. She can then pre-program the outcomes of the measurements in such a way that the outcomes are completely predictable ($G=1$), while allowing Alice and Bob to attain any value of $S$ up to its maximum value of 4. On the other hand, if Eve has no prior knowledge of what will be measured ($P=\frac{1}{4}$), Alice's and Bob's actions are not predetermined and hence, we say that they have complete ``free will". Any attempts to pre-program the outcomes of the measurements with complete predictability ($G=1$) will result in values $S\leq2$, familiar from the standard CHSH inequality. Finally, if Eve gives up any intention of extracting information ($G=\half$), then Alice and Bob could share an arbitrary no-signalling distribution, which will allow any $S\leq 4$. 

\begin{figure}[!t]
\begin{center}
\includegraphics[width=0.48\textwidth]{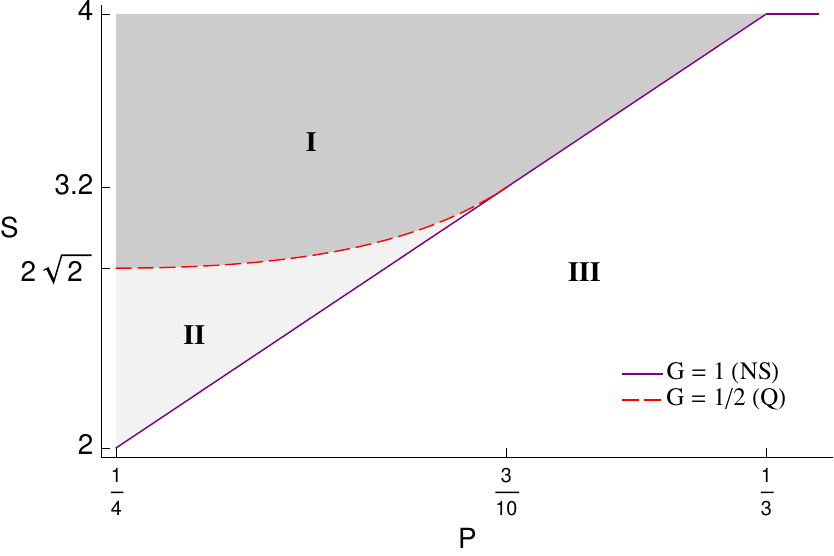} 
\vspace{-0.5cm} 
\caption{
(Color online) The maximal CHSH expectation value $S^{\max}(G,P)$ plotted against the free will parameter $P$, for the no-signalling (NS)  $G=1$ model (solid line), and the quantum (Q) $G=\half$ model (dashed line). Region III (unshaded) gives the set of $(S,P)$ values that can be attained by a deterministic $G=1$ model (purple line). This means if an $(S,P)$ value is found in its complement, i.e. Regions I (darker gray) and II (lighter gray), the model which corresponds to that point cannot be deterministic, i.e. Eve cannot know for certain what the outputs are. Regions II and III together give the set of $(S,P)$ values that may be obtained by a quantum $G=\half$ model (red line).
\label{SvsP}  }
\vspace{-0.5cm}
\end{center}
\end{figure}

\begin{figure}[!t]
\begin{center}
\includegraphics[width=0.48\textwidth]{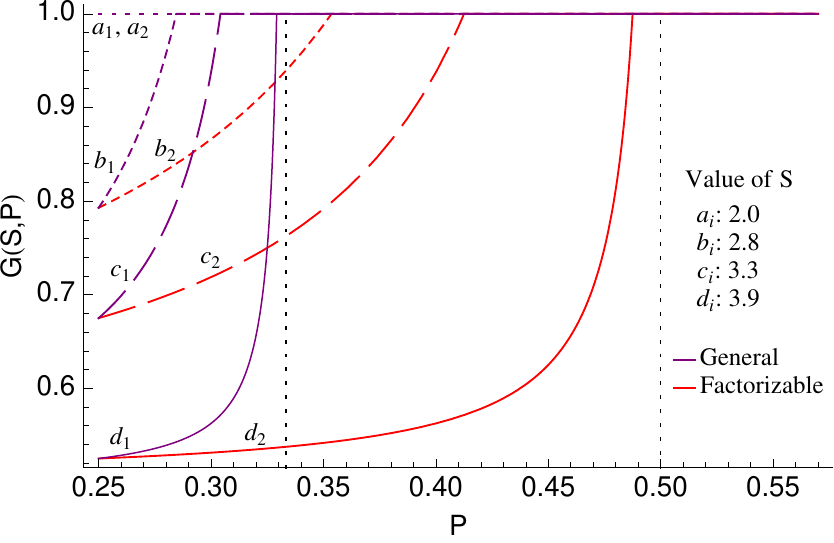}  
\vspace{-0.5cm}
\caption{(Color online) Optimal guessing probability $G (S ,P)$ for no-signalling models at different CHSH expectation values. $ S =2$ and $ S =2\sqrt{2}$ correspond to the local deterministic bound, and the Tsirelson bound respectively. For each pair of lines, the bound for the general case (purple) is always strictly greater than that of the factorizable case (red), except at $G=\half, 1$. In each case, the optimal guessing probabilities approach the vertical dotted lines as $S$ goes to $4$.
 }\label{GvsP}
\vspace{-0.5cm}
\end{center}
\end{figure}

From Theorem \ref{thm1}, we can estimate Eve's knowledge of Alice's and Bob's bits, quantified by $G$, given that we know the CHSH correlation $S$ that Alice and Bob observe as well as the free will parameter $P$. The bound in the theorem is tight, i.e.\ for any $G$ and $P$, there exists a no-signalling model for which the CHSH correlation is equal to $S^{\max}(G,P)$ (See Appendix \ref{sec:proof} for explicit constructions). In particular, suppose that Alice and Bob measure a CHSH correlation $S$. If $S \leq S^{\max}(1,P)$, then Alice and Bob know that the bits could have been completely pre-programmed before the Bell measurements were carried out. On the other hand, if $S > S^{\max}(1,P)$ (anywhere above the $G=1$ (NS) line in Fig.\ \ref{SvsP}), then Alice and Bob can conclude that some indeterminism has been introduced into the model, and that the guessing probability is less than unity. They can then use Eq.\ (\ref{eqthm1}) to determine an upper bound for the guessing probability $G$. For the case $P \geq \frac{1}{3}$, we have $S^{\max} (1,P) = 4$, which implies that $G=1$, i.e.\ Eve can use a deterministic protocol to achieve maximal Bell violation. The case where $P < \frac{1}{3}$ is more interesting because only in this case is the upper bound on the maximum guessing probability for a given CHSH correlation $S$ non-trivial:
\begin{equation}
G \leq \min \left\{\frac{1}{2} \left( 1 + \frac{4 - S}{4-S^{\max}(1,P)} \right), 1 \right\}, P < \frac{1}{3}. \label{generalG}
\end{equation}
The observed values for $S$ and $G$ thus give a tight upper bound on the guessing probability (Fig. \ref{GvsP}), from which the trade-off between the degree of free will and Bell violation can be seen.

Since our motivation is the task of randomness expansion, we need to evaluate the amount of true randomness that we can produce via post-processing. The degree to which this can be achieved is characterized by the min-entropy, which is used by a classical randomness extraction procedure in order to guarantee total privacy of a (shorter) random output string. For a single run, the min-entropy is defined to be $H_{\infty}(AB|XY)=-\log_{2}\max_{a,b,x,y}\tilde{p}(a,b|x,y)$, which is clearly bounded from below by $ -\log_{2} G$. For experimental estimation of a Bell violation, a Bell test must be performed on the devices many times in succession. This means that we must bound the min-entropy over a series of $n$ runs. If we assume that Eve can only perform a collective attack without memory, i.e.\ that the devices behave independently and identically at each run (as in the outlined model), then $\tilde{p}(r|s)=\tilde{p}(a^n b^n |x^n y^n)=\prod_{i} \tilde{p}(a_i b_i |x_i y_i)$ by independence and so $H_\infty (R|S)\geq -n \log_{2} G$.

\textit{Restricted adversary.---}So far, we have not placed any restrictions on the probability distribution $p(A_j, B_k|\lambda)$. Indeed, if Eve has quantum technology at her disposal, then she would be able to generate the most general of such distributions, i.e. RNGs share the entangled state $|\phi_\lambda\rangle = \sum_{j,k} \sqrt{p(A_j,B_k|\lambda)} |j\rangle\otimes |k\rangle$. In its absence, one should impose that the probability distributions are factorizable, i.e.\ $p(A_j, B_k|\lambda) = p^{(A)}(A_j|\lambda) p^{(B)}(B_k|\lambda)$. Note that while a non-factorizable distribution can always be made factorizable by utilizing more hidden variables, this changes the value of $P$. The generation of non-factorizable distributions requires a quantum state when the underlying variable $\lambda$ is not used. However this requires measurements in only one basis, hence in some sense, any `quantumness' cannot be detected. 

The results of Theorem \ref{thm1} hold in the case of an arbitrary probability distribution. Imposing the factorizability condition changes the upper bound for the Bell violation. In this case, as shown in Appendix \ref{sec:proof}, 

\begin{eqnarray}
S^{\max}_{\textrm{fac}} (G,P) = \begin{cases} 4 - 4(2G - 1)(1-2P) & P \leq \frac{1}{2}, \\ 4 & P \geq \frac{1}{2}.
\end{cases}   \label{eqthm2}
\end{eqnarray}
As expected, this does not exceed the bound in the more general case, implying that Eve has less influence in the factorizable case than in the general case. The upper bound on the guessing probability $G$ for an observed CHSH expectation value $S$ is analogous to Eq.\ (\ref{generalG}), upon replacing $S^{\max}$ with $S^{\max}_{\textrm{fac}}$ and the validity range by $P < \half$. Also, note that for $P = \frac{1}{4}$, corresponding to the case of complete free will, the bounds on $G$ for both the general and factorizable cases, reduce to the result in \cite{3}: $G \leq \frac{3}{2} - \frac{S}{4}$.

%For deterministic strategies, i.e. $G=1$, this gives $S(1,P) = \min\{24P-4,4\}$ and $S_{\textrm{fac}}(1,P) = \min\{8P,4\}$, representing a lower bound on any optimal indeterministic strategy (Fig. \ref{SvsP}). Any point above the graph must correspond to a $G<1$ strategy. In other words, Eve would be forced to introduce some indeterminism in the model, thus decreasing her guessing probability, if she wishes for a given $P$ to violate Bell inequality greater than $S(1,P)$ or $S_{\textrm{fac}}(1,P)$ respectively.

\textit{Quantum limit.---}The previously derived bounds apply under the weak assumption of no-signalling, which means that Eve might be able to distribute any no-signalling distribution between Alice and Bob, such as a PR box \cite{12, PRbox}, which can give the maximal violation of the CHSH inequality. However, if we make the stronger assumption of quantum mechanics, we know that Eve can achieve only much lower limits; in the case of $P=\frac{1}{4}$, she can do no better than $S=2\sqrt{2-(2G-1)^2}$ \cite{3}. In particular, for $G=\half$, the well-known Tsirelson bound $S=2\sqrt{2}$ is recovered. What happens when we move away from $P=\frac{1}{4}$? Imagine that Alice and Bob perform a CHSH test and calculate their expectation value averaged over all runs of the experiment. In addition, suppose that Eve uses a hidden variable model to determine the probabilities $p(A_j, B_k|\lambda)$ that, on a given run, Alice and Bob use to select their measurement settings. As far as Eve is concerned, she just has to optimize her quantum strategy for each of the different values of $\lambda$ independently, and the weighted probabilities $p(A_j, B_k|\lambda)$, which is effectively as if Alice and Bob (unbeknownst to them) are just playing a CHSH sub-game, with the correlation function given by
\begin{equation}
S(\lambda) = 4 \Big| \sum_{a,b,j,k } (-1)^{a+b+jk} \tilde{p}(a,b|A_j,B_k)p(A_j,B_k|\lambda)\Big|.
\end{equation}
In Appendix \ref{sec:quantum}, we derive the generalized Tsirelson bound for this class of games, and find the optimal distribution of probabilities to maximize $S(\lambda)$ for a given $P$. We also prove that for $P<\frac{3}{10}$, this maximum necessarily corresponds to the case $G=\half$. This implies that, for the optimal quantum strategy (meaning largest achievable CHSH expectation value), we have for $P< \frac{3}{10}$,
\begin{equation}
S^{\max}_Q (\half,P)= \frac{4(1-2P)^{3/2}} {\sqrt{(1-3P)}}. \label{quantumbound}
\end{equation}
For $P \geq \frac{3}{10}$, a deterministic strategy is used, and hence, $S^{\max}_Q (1,P)= S^{\max} (1,P)$.

%\begin{equation}
%S^{\max}_Q (G,P)= \begin{cases}
%\frac{4(1-2P)^{3/2}} {\sqrt{(1-3P)}} & G=\half, \, P<\frac{3}{10}, \\
%S^{\max} (1,P) & G=1, \, P\geq\frac{3}{10}, 
%\end{cases}
%\end{equation}
This considerably restricts the region of operation for Eve, as can be seen in Fig.\ \ref{SvsP}. Interestingly, for $P\geq\frac{3}{10}$, there is no quantum strategy that outperforms the deterministic strategy. This means that if Alice and Bob estimate that $P\geq\frac{3}{10}$, a randomness expansion protocol based on the CHSH inequality cannot function. We have not succeeded in finding a closed form for the general $S^{\max}(G,P)$ trade-off in the quantum strategy, except for recovering known limits such as $P=\frac{1}{4}$ \cite{3} and $G=1$ (Eq.\ (\ref{eqthm1})), although it can be solved numerically.

%\section{Concluding remarks}
\textit{Conclusions.---}We have shown that by influencing the apparently free choice of measurement settings in a Bell test, the adversary can fool the participants into thinking they share quantum correlations when, in fact, they do not and are being manipulated. We have specified the optimal models for Eve to maximize the guessing probabilities based on only no-signalling models, thereby specifying, for a given Bell correlation, a bound on the extent of private randomness that can be extracted. While this analysis has been performed assuming Eve has access to devices that can produce any no-signalling distribution (including PR boxes, etc.), we have compared this to the quantum case.

A natural extension of this work is to ask whether the local strategies employed here could be used to take advantage of a key distribution scheme, where Eve fakes a Bell violation to undermine the security that Alice and Bob believe is in their key. There are a number of subtleties that necessitate a more detailed study. %One could explicitly state a procedure that employs a Bell test for a subset of the total runs in $n$ experiments, yet generates a key from a disjoint subset, akin to the Ekert protocol \cite{9}.

This work is supported by the National Research Foundation and the Ministry of Education, Singapore. DEK is supported by Exploratory Initatives R-710-000-016-271. MJWH is supported by the ARC Centre of Excellence CE110001027. JEP acknowledges support from an EPSRC postgraduate studentship. CM is supported by EPSRC and the Istituto Superiore Mario Boella.

\newpage
\appendix

\section{Proof of Bell violation bounds and optimal models} \label{sec:proof}

In this appendix, we will prove Theorem \ref{thm1}, namely the tight relationship between the guessing probability, $G$, the free will parameter, $P$, and the observed CHSH expectation value, $S$, in both settings, where a generalized probability distribution is allowed (Eq.\ (\ref{eqthm1})), and where only a factorizable probability distribution is permitted (Eq.\ (\ref{eqthm2})). We start by defining $m_j = \tilde{p}^{(A)}(0|A_j,\lambda)$, $n_k = \tilde{p}^{(B)} (0|B_k,\lambda)$ for $j,k \in \{0,1\}$. Hence, 
$$
G(\lambda) = \max \{m_0, m_1, n_0, n_1, 1-m_0, 1-m_1, 1-n_0, 1-n_1 \}.
$$ 
If we also define $c_{jk} = \tilde{p} (0, 0|A_j, B_k, \lambda)$, then the other probabilities are readily expressed in these terms:
\begin{eqnarray*}
\tilde{p} (0, 1 |A_j, B_k, \lambda) &=& m_j - c_{jk},	\\
\tilde{p} (1,0 |A_j,B_k,\lambda) &=& n_k-c_{jk},			\\
\tilde{p} (1,1 |A_j, B_k,\lambda) &=& 1+ c_{jk} -m_j -n_k.
\end{eqnarray*}
In order to prove tight bounds, we will follow the techniques in \cite{7}. By Eq.\ (B3) of \cite{7}, 
$$
S \leq 4 - 2 \int d\lambda \  J(\lambda),
$$
where $J(\lambda) = \rho(\lambda|A_0, B_0)|m_0-n_0| + \rho(\lambda|A_0, B_1)|m_0-n_1| + \rho(\lambda| A_1, B_0)|m_1-n_0| + \rho(\lambda|A_1, B_1)|m_1+n_1 - 1|
$ and $\rho(\lambda|A_j,B_k)$ is the probability distribution of $\lambda$ given inputs $A_j, B_k$. An upper bound for $S$ corresponds to a lower bound for $J(\lambda)$. From the definition of $J(\lambda)$,
$J(\lambda) \geq (|m_0-n_0| + |m_0-n_1| + |m_1 - n_0| + |m_1+n_1 - 1|) \min_{j,k} \rho(\lambda|A_j, B_k)$. Consider the expression $K = |m_0-n_0| + |m_0-n_1| + |m_1 - n_0| + |m_1+n_1 - 1|$. By applying the triangle inequality to the first and second terms and to the third and fourth terms, we obtain $K \geq |n_0 - n_1| + |n_0 + n_1 - 1|$. 
Similarly, applying the triangle inequality to the first and third terms and to the second and fourth terms gives $K \geq |m_0 - m_1| + |m_0 + m_1 - 1|$.

Since $G(\lambda)$ was defined as the maximum of a set of 8 elements $\{m_0, m_1, n_0, n_1, 1-m_0, 1-m_1, 1-n_0, 1-n_1\}$, it has to be equal to at least one of them. Without loss of generality, suppose that $G(\lambda)=n_0$. Then $n_0 \geq n_1, 1 - n_1$, which implies that $K \geq n_0 - n_1 + n_0 + n_1 - 1 = 2G(\lambda)-1$. Consequently, $J(\lambda) \geq (2G(\lambda)-1) \min_{j,k} \rho(\lambda|A_j, B_k)$. By Bayes' Theorem and our assumption that $p(A_j, B_k) = \frac{1}{4}$, we obtain
\begin{equation}
S \leq 4 - 8 \int d\lambda \ (2G(\lambda)-1) \rho(\lambda) \min_{j,k} p(A_j, B_k|\lambda).
\end{equation}
We can now consider different allowable sets of $p$.

\begin{table}[!t]
    \begin{tabular}{c|c|c|c|c|c|c}
    $\lambda$  & $A_jB_k$ & $p_\lambda(A_jB_k)$ & $\tilde{p}_{jk\lambda}(00)$ & $\tilde{p}_{jk\lambda}(11)$ & $\tilde{p}_{jk\lambda}(01)$ & $\tilde{p}_{jk\lambda}(10)$ \\ \hline
     \multirow{4}{*}{$\lambda_1$} & $A_0B_0$ & $P$ & $G$ & $1-G$ & 0 &0 \\
    & $A_0B_1$ & $P$ & $G$ & $1-G$ & 0 &0 \\
    & $A_1B_0$ & $P$ & $G$ & $1-G$ & 0 &0 \\
    & $A_1B_1$ & $1-3P$ & $2G-1$ & 0 & $1-G$ & $1-G$\\ \hline
    \multirow{4}{*}{$\lambda_2$} & $A_0B_0$ & $P$ & $G$ & $1-G$ & 0 &0 \\
    & $A_0B_1$ & $P$ & $G$ & $1-G$ & 0 &0 \\
    & $A_1B_0$ & $1-3P$ & $1-G$ & $1-G$ & 0 &$2G-1$ \\
    & $A_1B_1$ & $P$ & 0 & 0 & $1-G$ & $G$\\ \hline
    \multirow{4}{*}{$\lambda_3$} & $A_0B_0$ & $P$ & $G$ & $1-G$ & 0 &0 \\
    & $A_0B_1$ & $1-3P$ & $1-G$ & $1-G$ & $2G-1$ &0 \\
    & $A_1B_0$ & $P$ & $G$ & $1-G$ & 0 &0 \\
    & $A_1B_1$ & $P$ & 0 & 0 & $G$ & $1-G$\\ \hline
    \multirow{4}{*}{$\lambda_4$} & $A_0B_0$ & $1-3P$ & $1-G$ & $1-G$ & $2G-1$ &0 \\
    & $A_0B_1$ & $P$ & $G$ & $1-G$ & 0 &0 \\
    & $A_1B_0$ & $P$ & $1-G$ & $G$ & 0 & 0 \\
    & $A_1B_1$ & $P$ & 0 & 0 & $1-G$ & $G$ 
    \end{tabular}
\caption{Optimal indeterministic model with guessing probability $G$ in the general case, for $\frac{1}{4} \leq P \leq \frac{1}{3}$. Notation:  $p_\lambda(A_j B_k)$ means $p(A_j, B_k| \lambda)$ and $\tilde{p}_{jk\lambda}(ab)$ means $\tilde{p}(a, b|A_j, B_k, \lambda)$. }
\label{qtable}
\end{table}

\textit{General case}: Using the definition of $P$ in Eq.\ (\ref{P}), for $P \geq \frac{1}{3}$, we could choose $p(A_j, B_k| \lambda) = (P, Q, Q',0)$ for some $Q, Q' \leq P$, in some order, for each $j,k \in \{0,1\}$. Then $\min_{j,k} p(A_j, B_k|\lambda) = 0$, from which it follows that $S^{\max}=4$. For $\frac{1}{4} \leq P < \frac{1}{3}$, given that $P$ is the largest probability, we have $\min_{j,k} p(A_j, B_k|\lambda)\geq 1-3P\geq 0$. This bound is sufficient to obtain the expression in Eq.\ (\ref{eqthm1}).

\textit{Factorizable case}: For a separable distribution with fixed $P=P_AP_B$, where $P_A = \max_{j,\lambda} p^{(A)}(A_j|\lambda)$ (and similarly for $P_B$), we need to bound the minimum probability, $(1-P_A)(1-P_B)$. This is equivalent to maximizing $P_A + P_B$ subject to the conditions $P_A P_B =P$ and $P_A ,P_B \leq 1$. The optimal values are found to be $P_A, P_B = (1, P)$ for $P \geq \frac{1}{2}$ and $(\frac{1}{2}, 2P)$ for $P \leq \frac{1}{2}$ in some order. These values give the bound in Eq.\ (\ref{eqthm2}).

In order to show that these bounds are tight, we have explicitly constructed a no-signalling model. For $P\leq \frac{1}{3}$ and a general probability distribution, this is given in Table \ref{qtable}. A $P\geq \frac{1}{3}$ model can be obtained in a similar way, for example, by replacing the column $p_\lambda(A_jB_k)$ of Table \ref{qtable} by the values, in the same order, $((P, Q, Q', 0)$, $(Q', P, 0, Q)$, $(Q, 0, P, Q')$, $(0, Q', Q, P))$, for any $Q, Q' \leq P$ that satisfies $Q + Q' + P = 1$. When $P \geq \frac{1}{2}$ and $Q' = 0$, we get an optimal model for the factorizable case. Replacing the column $p_\lambda(A_jB_k)$ of Table \ref{qtable} by the values, in the same order, $((P, P, \frac{1}{2} - P, \frac{1}{2} - P)$, $(\frac{1}{2} - P , P, \frac{1}{2} - P, P)$, $(P,\frac{1}{2} - P, P, \frac{1}{2} - P)$, $(\frac{1}{2} - P, \frac{1}{2} - P, P, P))$ gives an optimal $P \leq \frac{1}{2}$ model for factorizable distributions. 

\section{Optimal quantum strategies} \label{sec:quantum}

%Need to incorporate with main text
While we have demonstrated the optimal strategies when allowing for an arbitrary no-signalling distribution, it is equally interesting to apply the more physical restriction of assuming quantum mechanics. The question is, for a given $S$, how can (in the black-box scenario) Eve select the quantum state and influence Alice's and Bob's choices of measurement settings in order to maximize her guessing probability.

We start by asking a simpler question -- given a particular distribution of measurement settings $p_{jk} = p(A_j, B_k|\lambda)$, what is the maximum value of $S$ that Eve can possibly achieve? We will then show that for all these optimal strategies, we have $G=\half$, and that for a given $P=\max p_{jk}$, this value of $S$ is optimized when $\min p_{jk}=1-3P$ (for $P\geq 1/3$ the previous deterministic model suffices to achieve $S=4$). This therefore describes exactly the region that a quantum Eve can have a non-trivial guessing probability. The task of deriving a closed-form dependence of $G$ on $(S,P)$ between the $G=1$ deterministic line and $G=\half$ quantum line appears to be non-trivial except in special cases.

Consider a CHSH inequality with non-uniform measurement distribution. We maximize
\begin{equation*}
S_Q (\overline{p}) = 4 \ \Big| \sum_{j,k \in \{0,1\}} (-1)^{jk} \langle \psi| p_{jk} A_j B_k |\psi\rangle \Big| \end{equation*}
for a general probability distribution $\overline{p} =(p_{00},p_{01},p_{10},p_{11})$. For such a two-setting, two-outcome scenario, it is sufficient to consider just a pair of qubits \cite{qubitredux}. Hence, the maximization is restricted to all two-qubit states $\ket{\psi}$ and all local operators $A_j ,B_k$. In the black-box scenario, Alice and Bob interpret the outputs as $\pm1$ values, imposing that the $A_j$ and $B_k$ have these as their eigenvalues, i.e.\ are Pauli-like. This maximization was achieved in a restricted scenario in \cite{13}, which gave the maximum value of $S$ when the measurement distribution exhibits only local bias, i.e.\ Alice chooses 0 with probability $p$ and Bob chooses 0 with probability $q$ so that $\overline{p}=(pq,p(1-q),(1-p)q,(1-p)(1-q))$. We extend their methods to the more general case that can be implemented by Eve.

First, we formulate a Tsirelson--type bound on the expectation of outcome correlations:
\begin{eqnarray}
%\label{EQtwoparts}
\frac{S_Q (\overline{p})}{4} 
&\!\!\!\leq &\!\!  ||A_{0}\otimes (p_{00}B_{0} + p_{01}B_{1} ) |\psi \rangle||\! \nonumber\\ && \hspace{23 mm}+\,\,  ||A_{1}\otimes (p_{10}B_{0} - p_{11}B_{1} ) |\psi\rangle||\nonumber\\ &\!\!
\leq &\!\!\!\sqrt{ p_{00}^{2} + p_{01}^{2} + p_{00}p_{01}\alpha }\!+\!\sqrt{p_{10}^{2} +p_{11}^{2} - p_{10} p_{11} \alpha }, \nonumber\\ \label{EQbound}
\end{eqnarray}
where 
\begin{equation}
\alpha = \bra{\psi}I \otimes (B_{0}B_{1}+B_{1}B_{0})\ket{\psi}. \label{alpha}
\end{equation}
A maximum is achieved when \begin{equation} \label{thisisalpha}
\alpha = \frac{p_{00}^{2}p_{01}^{2}(p_{10}^{2}+p_{11}^{2})-p_{10}^{2}p_{11}^{2}(p_{00}^{2}+p_{01}^{2})}{p_{00}p_{01}p_{10}p_{11}(p_{00}p_{01}+p_{10}p_{11})}.
\end{equation}

Substituting Eq.\ (\ref{thisisalpha}) into Eq.\ (\ref{alpha}), we obtain
\begin{equation}
\label{symexpr} S_{Q}^{\max }(\overline{p} )= 4\sqrt{p_{00}p_{01}+p_{10}p_{11}}\sqrt{\frac{p_{00}^{2}+p_{01}^{2}}{p_{00}p_{01}}+\frac{p_{10}^{2}+p_{11}^{2}}{p_{10}p_{11}}}.
\end{equation}

While Eq.\ (\ref{alpha}) can be satisfied only if $|\alpha|\leq 2$, for $p_{jk}$ which satisfy
\begin{equation}
\label{condition} \frac{1}{p_{00}} +\frac{1}{p_{01}} +\frac{1}{p_{10}} +\frac{1}{p_{11}} -\frac{2}{p_{\min}} < 0 ,\end{equation} 
the expression for $\alpha$ in Eq.\ (\ref{thisisalpha}) gives $|\alpha|>2$. Hence, for $\alpha \geq 2$, the best approach is to use $\alpha = \pm 2$, for which the quantum bound coincides with a deterministic strategy $S_{D}^{\max}(\overline{p})=4-8p_{\min }$, where $p_{\min }$ is the smallest of the four probabilities (for example, if $p_{\min}=p_{11}$, this is achieved by pre-programming the devices to always output +1 regardless of the input). 

We now show that for all $\overline{p}$ with $-2 < \alpha < 2$ there exists a quantum strategy that achieves the quantum bound given in Eq. (\ref{symexpr}), which exceeds the deterministic bound $4-8 p_{\min}$.
We start by using freedom over local unitaries to specify that
\begin{eqnarray*}
B_0&=&X, \\
B_1&=&X\cos\beta+Z\sin\beta.
\end{eqnarray*}
However, due to the condition $\alpha = \bra{\psi}I \otimes (B_{0}B_{1}+B_{1}B_{0})\ket{\psi}$, this instantly imposes that $2\cos\beta=\alpha$. One can readily verify that by using the initial state and measurement settings
\begin{eqnarray}
\ket{\psi} &=& \frac{1}{\sqrt{2}}(\ket{00}+\ket{11}), \nonumber\\
A_{0} &=& \frac{(p_{00}+p_{01} \cos \beta )X+p_{01} \sin \beta Z}{\sqrt{(p_{00}+p_{01} \cos \beta )^{2}+(p_{01} \sin \beta)^{2}}},  \nonumber\\
A_{1} &=& \frac{(p_{10}-p_{11} \cos \beta )X-p_{11} \sin \beta Z}{\sqrt{(p_{10}-p_{11} \cos \beta )^{2}+(p_{11} \sin \beta)^{2}}}, \nonumber\\
B_{0} &=& X, \nonumber\\
B_{1} &=& X \cos \beta + Z \sin \beta, \label{settings}
\end{eqnarray}
the correct expectation value is realized. Note that $A_0$ and $A_1$ are the normalized versions of the operators $p_{00}B_0+p_{01}B_1$ and $p_{10}B_0-p_{11}B_1$, respectively. This choice was made in order for the equality to hold in Eq.\ (\ref{EQbound}).

Since $\ket{\psi}$ is a maximally entangled state, we know that the guessing probability $G=\half$. However, it remains to prove that this strategy is unique, up to local unitaries, i.e.\ that there isn't another strategy with a higher $G$. We already know that Bob's operators are uniquely specified up to local unitaries. In order to see that Alice's operators are also uniquely specified, it suffices to realize that in order to saturate the bound, it must be that whichever state $\ket{\psi}$ used must simultaneously be the maximal eigenvector of $A_{0}\otimes (p_{00}B_{0} + p_{01}B_{1} )$ and $A_{1}\otimes (p_{10}B_{0} - p_{11}B_{1} )$. Again, we use local unitary freedom, this time on Alice's side, to specify $A_0$. For instance, we could define $A_0$ as we did in Eq.\ (\ref{settings}), and $A_1=X\cos\gamma+Z\sin\gamma$. We can now diagonalize both operators and ascertain when they have a simultaneous maximal eigenvector. Up to sign changes such as $\gamma\mapsto-\gamma$ (which are associated with a further local unitary freedom), the unique result is that of the $A_1$ used previously. It is then easy to check that the overall operator has only one maximal eigenvector, which is $\ket{\psi}$. Hence, we can conclude that in the quantum limit (and within the quantum regime), $G=\half$. Of course, in the region $\alpha=2$ we know that $G=1$.

Finally, we investigate the measurement distribution $\overline{p}$ which yields the largest violation for a given $P= \max \{ p_{00},p_{01},p_{10},p_{11} \}$. The probabilities need not be ordered, and without loss of generality we choose the largest and smallest of these probabilities to be, say, $p_{00}=p_{\max}$ and $p_{11}=p_{\min}$. The normalization is
$$
p_{00} = 1-p_{01}-p_{10}-p_{11}.
$$
To find the optimal choice of $S_{Q}^{\max }$, we differentiate it with respect to $p_{01}$. This derivative is non-negative in the range $p_{01} \in [p_{\min}, p_{\max}]$, and zero if and only if $p_{01}=p_{\max}=P$, thus this choice maximizes $S_{Q}^{\max}$. By symmetry, we also demand that $p_{10}=P$. Therefore the optimal measurement distribution (up to permutations) is $(P,P,P,1-3P)$. Via Eq.\ (\ref{condition}), we see that such a quantum strategy gives an advantage over deterministic strategies only when $P<\frac{3}{10}$. When $\frac{3}{10} \leq P \leq \frac{1}{3}$, this distribution still yields the largest violation, but is achieved through a deterministic strategy with $S^{\max} (1,P) = 24P-4$, coinciding with the bound given in the main text for a general distribution with $G=1$.

We conclude that the optimal quantum strategy for a fixed $\frac{1}{4}\leq P<\frac{3}{10}$ and $G=\half$ gives the maximum CHSH expectation value stated in Eq.\ (\ref{quantumbound}).
Since Eve clearly has access to quantum technology, there is no reason to restrict the probability distribution $\overline{p}$. Nevertheless, one would obtain
$$
S^{\max}_{Q, \textrm{fac}}=4\sqrt{4P^2+(1-2P)^2},
$$
for any $\frac{1}{4}\leq P<\half$.


\begin{thebibliography}{99}
\bibitem{footnote} Indeed, in \cite{2}, the optimal way of tampering with free will leads to choosing three pairs of settings with equal probability $P$ and the last pair with probability $1-3P$; when $P=\frac{2+\sqrt{2}}{12}$, singlet statistics are recovered. This can be reinterpreted as a convex combination: in a fraction $f$ of cases, Alice and Bob have full free will, so each pair of settings is chosen with probability $\frac{1}{4}$; in the remaining $1-f$ cases, a device is used, such that the last pair of settings is never chosen. By identification, free will must be given up in a fraction of runs $1-f=\sqrt{2}-1\approx 41\%$.
\end{thebibliography}
\end{document}